\documentclass[twocolumn,superscriptaddress,floatfix,showpacs]{revtex4-1}
\usepackage{amsmath,amsfonts,amssymb}
\usepackage[pdftex]{color,graphicx,hyperref}
\pdfoutput=1
\usepackage{subfigure}
\usepackage{graphicx}
\usepackage{multirow}
\usepackage{array}
\usepackage{colortbl}
\begin{document}

\def\be{\begin{equation}}
\def\ee{\end{equation}}
\def\bc{\begin{center}}
\def\ec{\end{center}}
\def\bea{\begin{eqnarray}}
\def\eea{\end{eqnarray}}
\newcommand{\avg}[1]{\langle{#1}\rangle}
\newcommand{\Avg}[1]{\left\langle{#1}\right\rangle}

\title{Entropy of dynamical social networks}

\author{Kun Zhao} 
\affiliation{Physics Department, Northeastern University, Boston, Massachusetts,  United States of America}
\author{M\'arton Karsai}
\affiliation{BECS, School of Science, Aalto University, Aalto, Finland}
\author{Ginestra Bianconi}
\email{ginestra.bianconi@gmail.com}
\affiliation{Physics Department, Northeastern University, Boston, Massachusetts,  United States of America}

\date{\today}

\begin{abstract}
Human dynamical social networks encode information and are  highly adaptive. To characterize the information encoded in the fast dynamics of social interactions, here we introduce the entropy of  dynamical social networks. By analysing a large  dataset  of phone-call interactions we show evidence that the dynamical social network has an entropy that depends on the time of the day in a typical week-day.  Moreover  we show evidence for adaptability of human social behavior showing  data on duration of phone-call interactions that significantly deviates from the statistics of duration of face-to-face interactions. This adaptability of behavior corresponds to a different information content of the dynamics of social human interactions. We  quantify this information by the use of the entropy of dynamical networks on realistic models of social interactions. \footnote{Published in PLoS ONE 6(12): e28116 (2011).}
\end{abstract}

\maketitle

\section*{Introduction}
Networks \cite{Dorogovtsev:2003,Newman:2003,Boccaletti:2006,Caldarelli:2007,Barrat:2008} encode information in the topology of their  interactions. This is the main reason why networks are ubiquitous in complexity theory and constitute the underlying structures of  social, technological and biological systems. The information encoded in social networks \cite{Granovetter:1973, Wasserman:1994} is  essential to build strong collaborations \cite{Newman:2001} that   enhance the performance of a society, to build reputation trust and to navigate \cite{Kleinberg} efficiently the networks. For these reasons social networks are small world \cite{WS} with short average distance between the nodes  but large clustering coefficient.
Therefore to understand how social network evolve, adapt and respond to external stimuli, we need to develop a  new information theory of complex social networks.

Recently, attention has been addressed to entropy measures applied to email correspondence \cite{Eckmann:2004},  static networks \cite{Bianconi:2008,Bianconi:2008b,Anand:2009} and mobility patterns \cite{Song:2010}. 
New network entropy measures quantify the information encoded in  heterogenous  static networks \cite{Bianconi:2008,Anand:2009}. Information theory tools set the limit of predictability of human mobility\cite{Song:2010}.  Still we lack methods to assess the information encoded in the dynamical social interaction networks.

Social networks are characterized by complex organizational structures revealed by network community  and degree correlations  \cite{Fortunato}. These structures  are sometimes correlated with annotated features of the nodes or of the links such as age, gender, and other annotated features  of  the links such as shared interests, {family} ties or common  work locations  \cite{Palla:2007,Lehmann}. In a recent work  \cite{Bianconi:2009} it has been shown by studying social, technological and biological networks that the network entropy measures can assess how significant are the annotated features for the network structure.

Moreover social networks evolve on many different time-scales and relevant information is encoded in their dynamics.  In fact social networks are highly  adaptive.
Indeed social ties can appear or disappear depending on  the dynamical process occurring on the networks such as epidemic spreading or opinion dynamics.
Several models for adaptive social evolution have been proposed showing phase transitions in different universality classes
 \cite{Bornholdt:2002,Marsili:2004,Holme:2006,MaxiSanMiguel:2008}.
Social ties have in addition to that a microscopic structure constituted by  fast social interactions of the duration of a phone call or of a face-to-face interaction.  
Dynamical social networks characterize the social interaction at this fast time scale. For these dynamical networks new network  measures are starting to be defined  \cite{Latora:2009} and recent works focus on the implication that the network dynamics has on percolation, epidemic spreading and opinion dynamics  \cite{Holme:2005,Vazquez:2007,Havlin:2009,Isella:2011,Karsai:2011a}.

Thanks  to the availability of new extensive data on a wide variety of human dynamics \cite{Barabasi:2005,Eagle:2006,Rybski:2009,Amaral:2008,Amaral:2009}, human mobility  \cite{Brockmann:2006,Gonzalez:2008,Song:2010} and dynamical social networks  \cite{Onnela:2007}, it has been recently recognized that many human activities  \cite{Vazquez:2007}  are bursty and not Poissonian.  New data on social dynamical networks  start to be collected with new technologies such as of Radio frequency Identification Devices \cite{Cattuto:2010,Isella:2011} and Bluetooth \cite{Eagle:2006}. These technologies are able to record the duration of  social interactions and report evidence for a bursty nature of social interaction characterized by a fat tail distribution of {the duration of} face-to face interactions. This bursty behavior of social networks  \cite{Hui:2005,Scherrer:2008,Cattuto:2010,Isella:2011,Stehle:2010,Zhao:2011} is coexisting with modulations coming from periodic daily (circadian rhythms) or weakly patterns  \cite{Karsai:2011b}. The fact that this bursty behavior is observed also in social interaction of simple animals, in the motion of rodents  \cite{Chialvo}, or in the use of words  \cite{Motter}, suggests that the underlying origin of this behavior is  dictated by the biological and neurological processes underlying the dynamics of the social interaction. To our opinion this problem remains open:
How much can humans intentionally change the statistics of social interactions and the level of information encoded in the dynamics of their social networks, when they are interfacing with a new technology?

In this paper we try to address this question by studying the dynamics of {interactions through phone calls} and comparing it with face-to-face interactions. We show that the entropy of dynamical networks is able to quantify the information encoded in the dynamics of phone-call interactions during a typical week-day.
Moreover we show evidence that human social behavior is highly adaptive and that the duration of face-to-face interaction in a conference follows a different distribution than duration of phone-calls. We therefore have evidence of an intentional capability of humans to change statistically their behavior when interfacing with the technology of mobile phone communication. Finally we  develop a model in order to quantify how much the entropy of dynamical networks changes if we allow modifications in the distribution of duration of the interactions.

\section*{Results}

\subsection*{Entropy of dynamical social networks}
\begin{figure}[!ht]
\begin{center}
\includegraphics[width=3.27in]{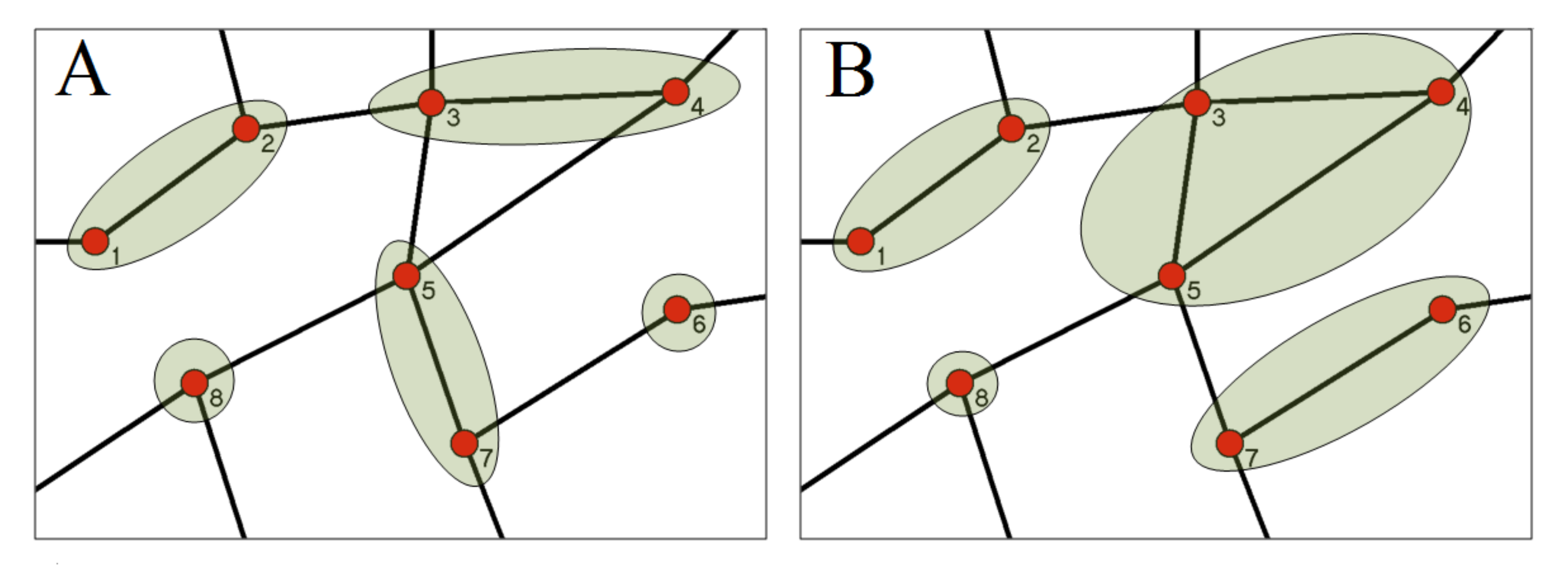}
\end{center}
\caption{  {\bf The dynamical social networks are composed by different dynamically changing groups of interacting agents.} In panel (A) we allow only for groups of size one or two as it typically happens in mobile phone communication. In panel (B) we allow for groups of any size as in face-to-face interactions. }
\label{fig1}
\end{figure}

In this section we introduce the entropy of dynamical social networks as a measure of information encoded in their dynamics. Since we are interested in the dynamics of contacts we assume to have a quenched social network $G$ of friendships, collaborations or acquaintances formed by $N$ agents and we allow  a dynamics of social interactions on this network.
If two agents $i,j$ are linked in the network they can meet and interact at each given time giving rise to the dynamical social  network under study in this paper.
If a set of agents of size $ N$ is connected through the social network $G$ the agents $i_1,i_2,\ldots i_n$ can interact in a group of size $n$.
Therefore at any given time the static network $G$ will be partitioned in connected components or groups of interacting agents as shown in Fig \ref{fig1}.
In order to indicate that a social interaction is occurring at time $t$ in the group of agents $i_1,i_2,\ldots , i_n$ and that these agents are not interacting with other agents, we write $g_{i_1,i_2,\ldots , i_n}(t)=1$ otherwise we put $g_{i_1,i_2,\ldots, i_n}(t)=0$.
Therefore each agent is interacting with one group of size $n>1$ or non interacting (interacting with a group of size $n=1$).
Therefore at any given time
\begin{equation}
\sum_{{\cal G}=(i,i_2,\ldots,i_n )|i\in {\cal G}}g_{i,i_2,\ldots, i_n}(t)=1.
\end{equation} 
where we  indicate with ${\cal G}$ an arbitrary connected subgraph of $G$.
The history ${\cal S}_t$ of the dynamical social network is given by ${\cal S}_t=\{g_{i_1,i_2,\ldots, i_n}(t^{\prime})\, \forall t^{\prime}<t\}$.
If we indicated by $p(g_{i_1,i_2,\ldots, i_n}(t)=1|{\cal S}_t)$ the probability that $g_{i_1,i_2,\ldots, i_n}(t)=1$ given the story ${\cal S}_t$, 
the likelihood that at time $t$ the dynamical networks has a group configuration $g_{i_1,i_2,\ldots,i_n}(t)$ is given by 
\begin{equation}
{\cal L}=\prod_{{\cal G}} p(g_{i_1,i_2,\ldots, i_n}(t)=1|{\cal S}_t)^{g_{i_1,i_2,\ldots, i_{n}}(t)} 
\end{equation}

The entropy $S$ characterizes the logarithm of the typical number of different group configurations that can be expected in the dynamical network model at time $t$ and is given by $S=-\Avg{\log{\cal L}}_{|{\cal S}_t}$ that we can explicitly express as
\begin{equation}
S=- \sum_{{\cal G}}p(g_{i_1,i_2,\ldots, i_n}(t)=1|{\cal S}_t)\log p(g_{i_1,i_2,\ldots, i_n}(t)=1|{\cal S}_t).
\end{equation}
According to the information theory results \cite{Cover:2006}, if the entropy is vanishing, i.e. $S=0$ the network dynamics is regular and perfectly predictable, if the entropy is larger the number of future possible configurations is growing and the system is less predictable. 
If we model face-to-face interactions we have to allow the  possible formation of groups of any size, on the contrary, if we model the mobile phone communication,  we need to allow only for pairwise interactions.
Therefore, if we define the adjacency matrix of the network $G$ as the matrix $a_{ij}$,  the log likelihood takes the very simple expression given by 
\begin{equation}
{\cal L}=\prod_i p(g_i(t)=1|{\cal S}_t)^{g_i(t)}\prod_{ij|a_{ij}=1} p(g_{ij}(t)=1|{\cal S}_t)^{g_{ij}(t)} 
\end{equation}
with 
\begin{equation}
g_i(t)+\sum_{j} a_{ij} g_{ij}(t)=1,
\end{equation}
for every time $t$.
The entropy is then given by 
\begin{eqnarray}
S&=&- \sum_i p(g_{i}(t)=1|{\cal S}_t)\log p(g_{i}(t)=1|{\cal S}_t)\nonumber \\
&&-\sum_{ij}a_{ij} p(g_{ij}(t)=1|{\cal S}_t)\log p(g_{ij}(t)=1|{\cal S}_t).
\end{eqnarray}

\subsection*{Social dynamics and entropy of phone call interactions}

We have analyzed the call sequence of subscribers of a major euroepan mobile service provider. We considered calls between users who at least once called each other during the examined $6$ months period in order to examine calls only reflecting trusted social interactions. The resulted event list consists of $633,986,311$ calls between $6,243,322$ users. For the entropy calculation we selected $562,337$ users who executed at least one call per a day during a week period. First of all we have  studied  how the entropy of this dynamical network is affected by circadian rhythms.
 We assign to each agent $i=1,2$ a
 number $n_i=1,2$ indicating the size of the group where he/she belongs. If an agent $i$
has coordination number $n_i=1$ he/she is isolated, and if 
$n_i=2$ he/she is interacting with a group of $n=2$ agents.
We also assign to each agent $i$ the variable $t_i$ indicating
the last time at which the coordination number $n_i$ has changed.
If we neglect the feature of the nodes, the most simple transition probabilities that includes for some memory effects present in the data,  is given by a probability $p_n=p_n(\tau,t)$ for an agent in state $n$ at time $t$  to change his/her state given that he has been in his/her current state for a duration  $\tau=t-t_i$.

\begin{figure}[!ht]
\begin{center}
\includegraphics[width=3.27in]{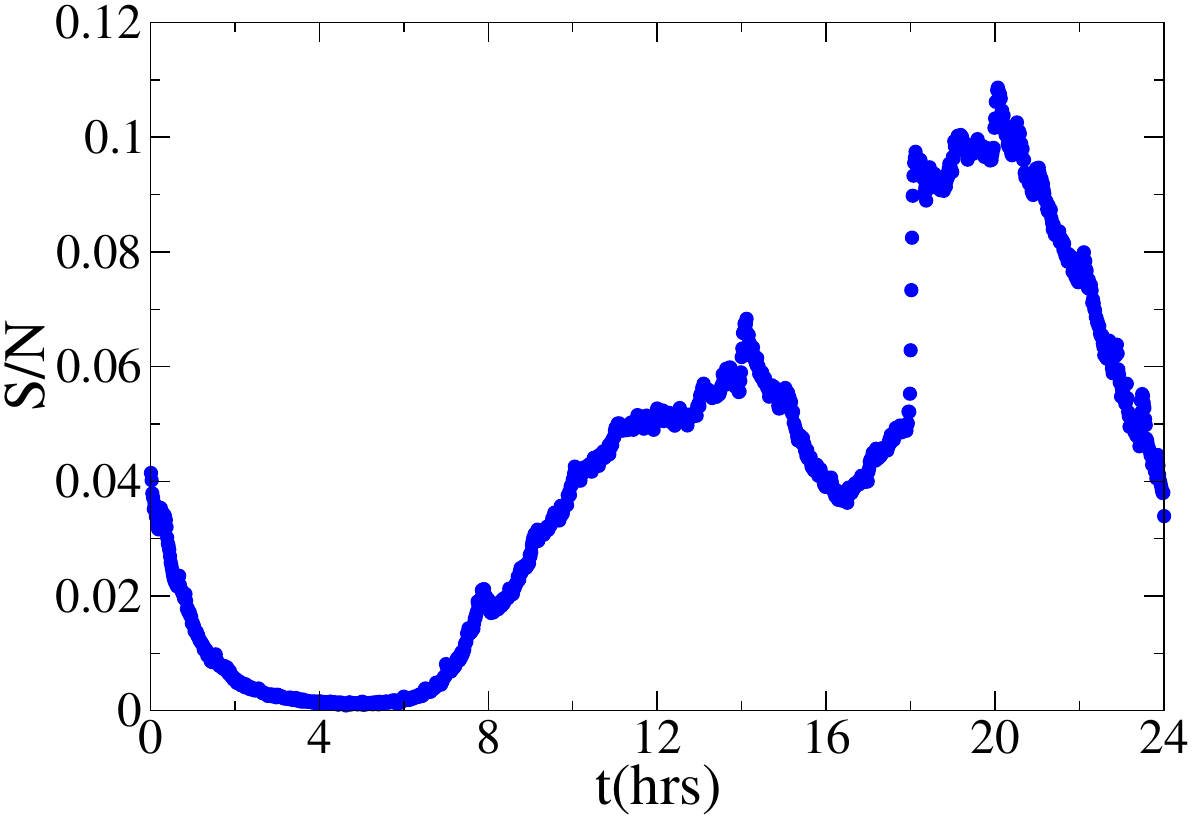}
\end{center}
\caption{  {\bf Mean-field evaluation of the entropy of the dynamical social networks of phone calls communication in a typical week-day.} In the nights the social dynamical network is more predictable.}
\label{entropyt}
\end{figure}

We have estimated the probability $p_n(\tau,t)$ in a typical week-day. Using the data on the probabilities $p_n(\tau,t)$  we have calculated the entropy, estimated by a mean-field evaluation (Check Text S1) of   the dynamical network as a function of time in a typical week-day. The entropy of the dynamical social network is reported in Fig. \ref{entropyt}. It significantly changes during the day  describing the fact that  the predictability of the phone-call networks change as a function of time. In fact, as if the  entropy of the dynamical network is smaller and the network is an a more predictable state.

\subsection*{Adaptive dynamics face-to face interactions and phone call durations}

In this section we report evidence of adaptive human behavior by showing that the duration of phone calls, a binary social interactions mediated by technology, show different statistical features respect to face-to-face interactions.
The distributions of the
times describing human activities are typically broad
 \cite{Barabasi:2005,Vazquez:2007,Hui:2005,Rybski:2009,Cattuto:2010, Isella:2011}, and are closer to power-laws, which lack a
characteristic time scale, than to exponentials.
In particular  in  \cite{Cattuto:2010} there is  reported data on Radio Frequency Identification devices, with temporal resolution of 20s, showing that both distribution  duration of face-to-face contacts and inter-contact periods is fat tailed during  conference venues.

\begin{figure}[!ht]
\begin{center}
\includegraphics[width=70mm, height=65mm]{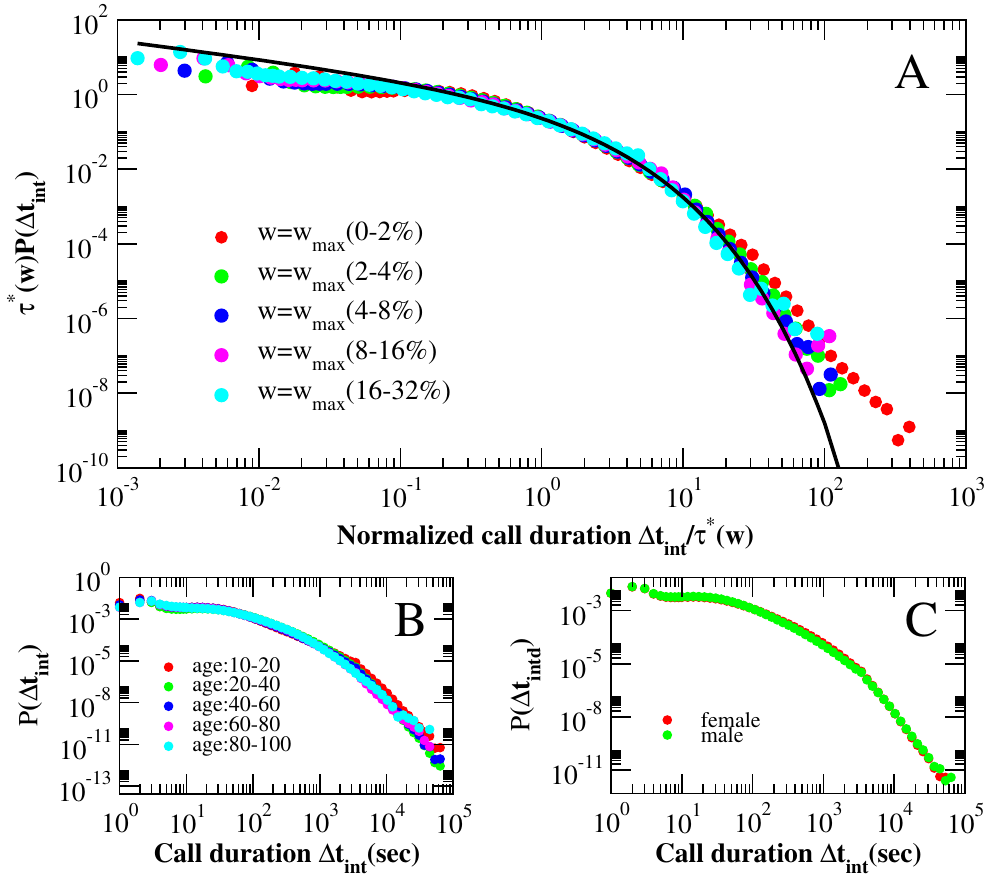}
\end{center}
\caption{ {\bf Probability distribution of duration of phone-calls.} (A) Probability distribution of duration of phone-calls between two given persons connected by a link of weight $w$. The data depend on the typical scale $\tau^{\star}(w)$ of duration of the phone-call.
(B) Probability distribution of duration of phone calls for people of different age. (C) Probability distribution of duration of phone-calls for people of different gender. The distributions shown in the panel (B) and (C) do not significantly depend on the attributes of the nodes.}
\label{interaction}
\end{figure}

\begin{table}[!ht]
\caption{\bf{}Typical times $\tau^{\star}(w)$ used in the data collapse of Fig. \ref{interaction}.}
\begin{tabular}{|c|c|}
\hline
Weight of the link & Typical time $\tau^{\star}(w)$ in seconds (s)\cr
\hline
(0-2\%) $w_{max}$ &111.6 \\
\hline
(2-4\%) $w_{max}$ & 237.8 \\
\hline
(4-8\%) $w_{max}$ & 334.4 \\
\hline
(8-16\%) $w_{max}$  & 492.0 \\
\hline
(16-32\%) $w_{max}$ & 718.8 \\
\hline
\end{tabular}
\begin{flushleft} 
\end{flushleft}
\label{tauw}
\end{table}

\begin{figure}[!ht]
\begin{center}
\includegraphics[width=70mm, height=60mm]{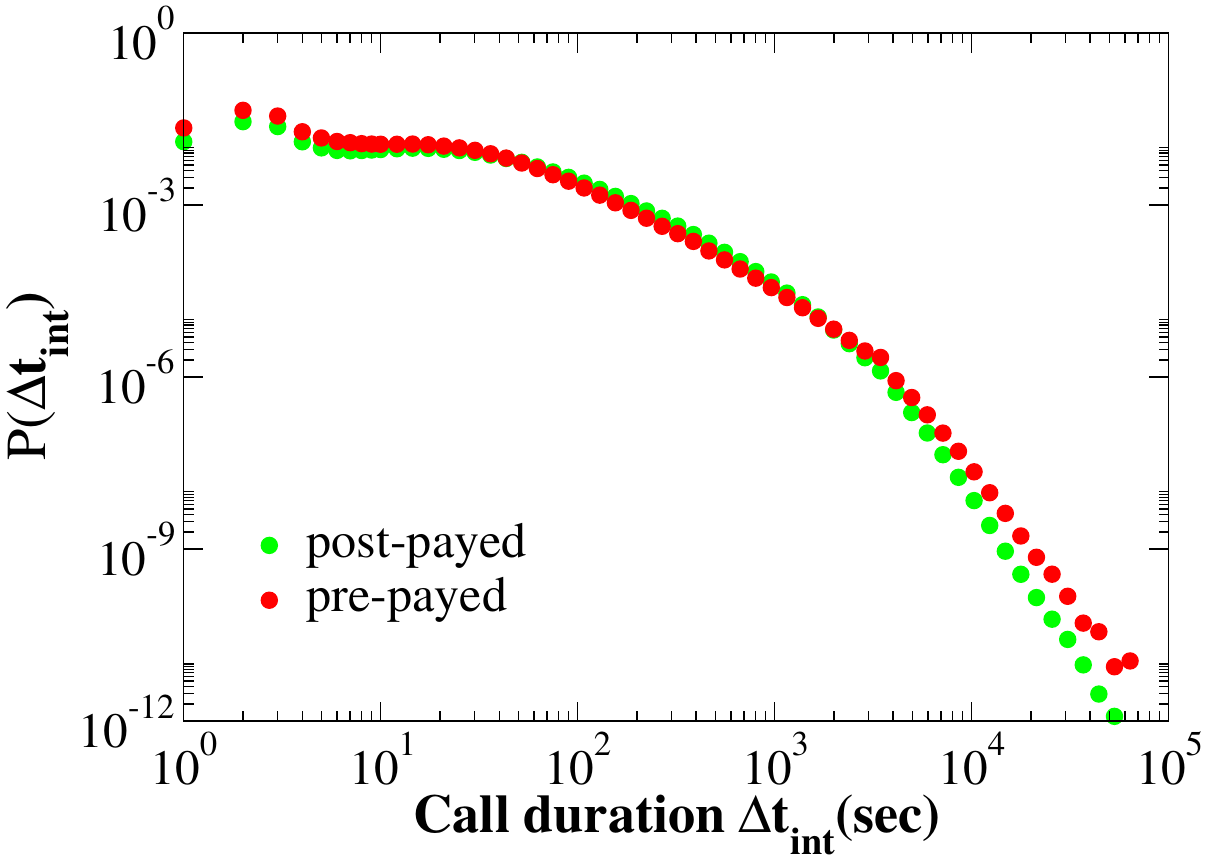}
\end{center}
\caption{  {\bf Probability distribution of duration of phone-calls for people with different types of contract.} No significant change is observed that modifies the functional form of the distribution.}
\label{pay}
\end{figure}

Here we analysed the above defined mobile-call event sequence performing the measurements on all the users for the entire 6 months time period. The distribution of phone-call durations strongly deviates from a fat-tail distribution.
In Fig. \ref{interaction} we report this distributions and show that these distributions  depend on the strength $w$ of the interactions (total duration of contacts in the  observed period) but do not depend on the age, gender or type of contract in a significant way.
The distribution $P^w(\Delta t_{in})$ of duration of contacts within agents with strenght $w$ is well fitted by a Weibull distribution
\begin{equation}
\tau^*(w) P^w(\Delta t_{in})=W_{\beta}\left(x=\frac{\Delta t}{\tau^{\star}(w)}\right)= \frac{1}{x^{\beta}} e^{-\frac{1}{1-\beta}x^{1-\beta}}.
\end{equation}
with $\beta=0.47..$.
{The typical times $\tau^*(w)$ used for the data collapse of Figure 3 are listed in Table \ref{tauw}}. The origin  of this  significant change in behavior of humans interactions could be due to the consideration of the cost of the interactions (although we are not in the position to draw these conclusions (See Fig. \ref{pay} in which we compare distribution of duration of calls for people with different type of contract)  or might depend on the different nature of the communication.
The duration of a phone call is quite short and is not affected significantly by the circadian rhythms of the population.
On the contrary the duration of no-interaction periods is strongly affected by periodic daily of weekly rhythms.
The distribution of no-interaction periods can be fitted by a double power-law but also a single Weibull distribution can give a first approximation to describe $P(\Delta t_{no}).$
In Fig. \ref{noninteraction} we report the distribution  of duration of no-interaction periods in the day periods between 7AM and 2AM next day. { The typical times $\tau^*(k)$ used in Figure 5  are listed in Table \ref{tauk}.} 

\section*{Discussion}
\subsection*{The entropy of a realistic model of cell-phone interactions}
The data on face-to-face and mobile-phone  interactions show that a reinforcement dynamics is taking place during the human social interaction. 
Disregarding for the moment the effects of circadian rhythms and weakly patterns, a possible
explanation of such results is given by mechanisms in which the
decisions of the agents to form or leave a group are driven by memory
effects dictated by reinforcement dynamics, that can be summarized in
the following statements: {\em i) the longer an agent is interacting in a
  group the smaller is the probability that he/she will leave the group;
 ii) the longer an agent is isolated the smaller is the probability that
  he/she will form a new group}. In particular, such reinforcement
principle implies that the probabilities $p_n(\tau,t)$ that an agent
with coordination number $n$ changes his/her state depends on the time
elapsed since his/her last change of state, i.e.,
$p_n(\tau,t)=f_n(\tau)$. 
To ensure the reinforcement dynamics  any function $f_n(\tau)$ which is a decreasing function of its argument can be taken.
In two recent papers\cite{Stehle:2010, Zhao:2011}  the face-to-face interactions   have been realistically modelled with the use of the reinforcement dynamics, by choosing
\begin{equation}
f_n(\tau)=\frac{b_n}{(\tau+1)}.
\end{equation}
 with good agreement with the data when we took  $b_n=b_2$ for $n\geq 2$ and $b_1>0$, $b_2>0$.

\begin{figure*}[!ht]
\begin{center}
\includegraphics[width=150mm, height=50mm]{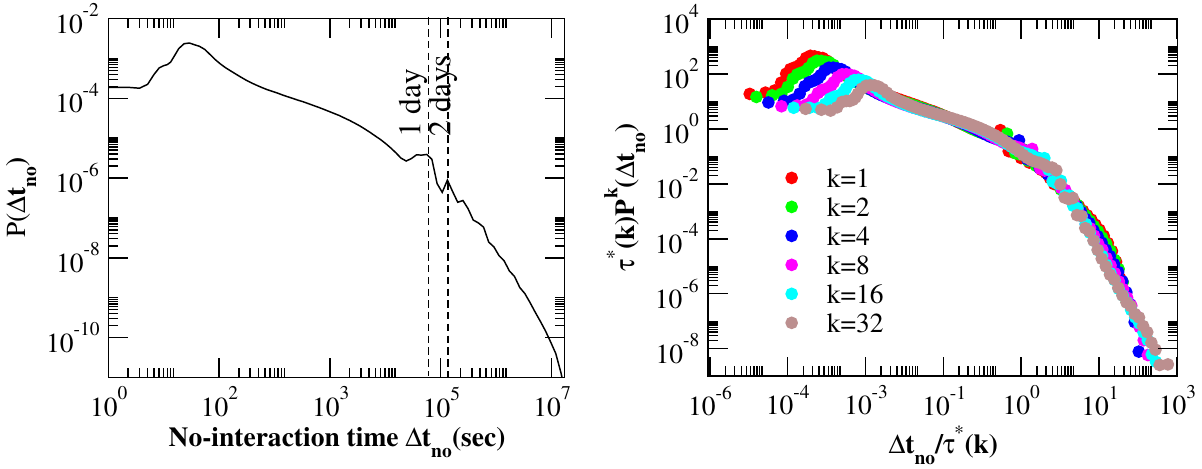}
\end{center}
\caption{ {\bf Distribution of non-interaction times in the phone-call data.} The distribution strongly depends on circadian rhythms. The distribution of rescaled time depends strongly on the connectivity of each node. Nodes with higher connectivity $k$ are typically non-interacting for a shorter  typical time scale $\tau^{\star}(k)$.}
\label{noninteraction}
\end{figure*}

In order to model the phone-call data studied in this paper we can always adopt the reinforcement dynamics but we need to modify the probability $f_n(\tau)$ by a  parametrization  with an additional parameter $\beta\leq1$.  In order to be specific in our model of mobile-phone communication, we consider a system that consists of $N$ agents. Corresponding to the mechanism of daily cellphone communication, the agents can call each other to form a binary interaction if they are neighbor in the social network.  The social network is characterized by a   given degree distribution  $p(k)$ and a given weight distribution $p(w)$.  Each agent $i$ is characterized by the size $n_i=1,2$ of the group he/she belongs to and the last time $t_i$ he/she has changed his/her state.
Starting from random initial conditions, at each timestep $\delta t=1/N$  we  take a random agent.
If the agent is isolated he/she will change his/her state with probability 
\begin{equation}
f_1^{\beta}(\tau)=\frac{b_1}{(\tau+1)^{\beta}}
\end{equation}
with $\tau=t-t_i$ and $b_1>0$.
If he/she change his/her state he/she will call one of his/her  neighbor in the social network {which is still not} engaged in a telephone call.
A non-interacting  neighbor agent will pick up the phone with probability $f_1^{\beta}(\tau')$ where $\tau'$ is the  time he/she has not been interacting.

\begin{table}[!ht]
\caption{\bf{Typical times $\tau^{\star}(k)$ used in the data collapse of Fig. \ref{noninteraction}.}}
\begin{tabular}{|c|c|}
\hline
Connectivity & Typical time $\tau^{\star}(k)$ in seconds (s) \\
\hline
k=1 &158,594 \\
\hline
k=2 &118,047 \\
\hline
k=4 & 69,741 \\
\hline
k=8 & 39,082 \\
\hline
k=16 & 22,824 \\
\hline
k=32 & 13,451 \\
\hline
\end{tabular}
\label{tauk}
\end{table}

If, on the contrary the agent $i$ is interacting, he/she  will change his/her state with probability $f_2^{\beta}(\tau|w)$ depending on the weight of the link and on the duration of the phone call.
We will take in particular
\begin{equation}
f_2^{\beta}(\tau|w)=\frac{b_2 g(w)}{(\tau+1)^{\beta}}
\end{equation}
where $b_2>0$ and $g(w)$ is a decreasing function of the weight $w$ of the link. The distributions $f_1^{\beta}(\tau)$ and $f_2^{\beta}(\tau|w)$ are parametrized by the parameter $\beta\leq1$. As $\beta$ increases, the distribution of duration of contacts and duration of intercontact time become broader.
These probabilities give rise to either Weibull distribution of duration of interactions (if $\beta<1$) or power-law distribution of duration of interaction $\beta=1$.
 Indeed for $\beta<1$, the probability $P^w_2(\tau)$ that a conversation  between two nodes with link weight $w$ ends after a duration $\tau $ is given by the Weibull distribution (See Text S1 for the details of the  derivation)
\begin{equation}
\tau^*(w) P^w_2(\tau)\propto W_{\beta}((\tau+1)/\tau^{\star}(w))
\end{equation}
with $\tau^{\star}(w)=\left[{2 b_2 g(w)}\right]^{-1/(1-\beta)}$. This distribution well capture  the distribution observed in mobile phone data and reported in Fig. \ref{interaction}
(for a discussion of the validity of the annealed approximation for predictions on a quenched network see the  Text S1 ).

If, instead of having $\beta<1$ we have $\beta=1$ the probability distribution for duration of contacts is given by a power-law 
\begin{equation}
P_2^w(\tau)\propto(\tau+1)^{-[2b_2g(w)+1]}.
\end{equation}
This distribution is comparable  with the distribution observed in face-to-face interaction during conference venues \cite{Stehle:2010, Zhao:2011}.
The adaptability of human behavior, evident when comparing the distribution of duration of phone-calls with the duration of face-to-face interactions, can be understood as a possibility to change the exponent $\beta$ regulating the duration of social interactions.

\begin{figure}[!ht]
\begin{center}
\includegraphics[width=3.27in ,height=50mm]{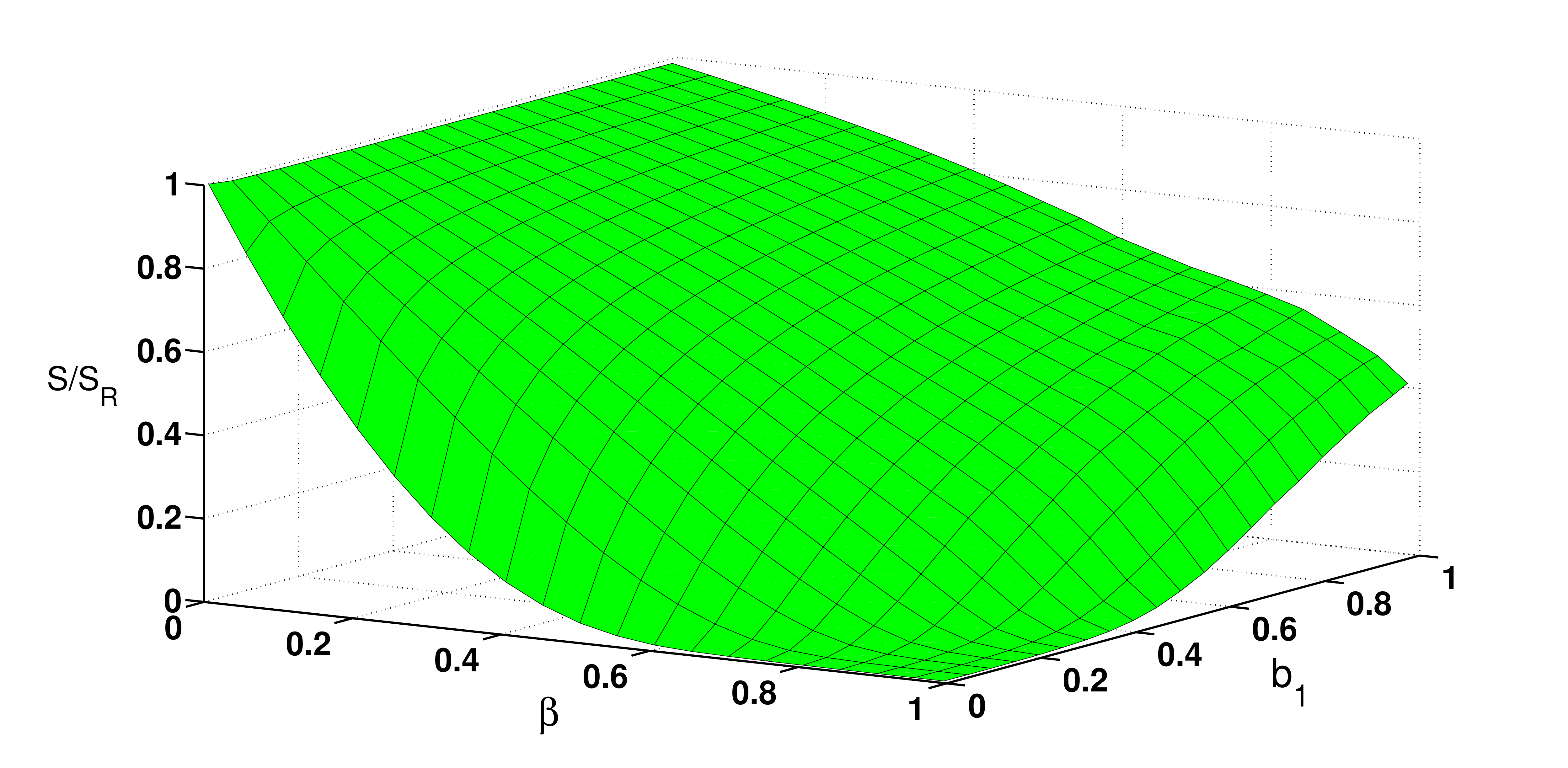}
\end{center}
\caption{  {\bf Entropy $S$ of social dynamical network model of pairwise communication normalized with the entropy $S_R$ of a null model in which the expected average duration of phone-calls is the same but the distribution of duration of phone-calls and non-interaction time are Poisson distributed.} The network size is $N=2000$ the degree distribution of the network is exponential with average $\avg{k}=6$, the weight distribution is $p(w)=Cw^{-2}$ and $g(w)$ is taken to be $g(w)=b_2/w$ with $b_2=0.05$.
The value of $S/S_R$ is depending on the two parameters $\beta, b_1$. For every value of $b_1$ the normalized entropy is smaller for $\beta\to 1$.
}
\label{entropy_network}
\end{figure}

Changes in the parameter $\beta$ correspond to a different entropy of the dynamical social network.
Solving analytically this model we are able to evaluate  the dynamical entropy  as a function of $\beta$ and $b_1$. 
In Fig.  \ref{entropy_network} we report the entropy $S$ of the dynamical social network  a function of $\beta$ and $b_1$ in the annealed approximation and the large network limit. In particular we have taken  a  network of size $N=2000$ with exponential degree distribution of average degree $\avg{k}=6$, weight distribution $P(w)=Cw^{-2}$ and function $g(w)=1/w$ and  $b_2=0.05$.  Our aim in   Fig. \ref{entropy_network} is to show  only the effects on the entropy due to the different distributions of duration of contacts and non-interaction periods. Therefore we have normalized the entropy $S$ with the entropy $S_R$ of  a null model of social interactions in which the duration of groups are Poisson distributed but the average time of interaction and non interaction time are the same as in the model of cell-phone communication. 
From Fig. \ref{entropy_network}  we observe that if we keep $b_1$ constant,  the ratio $S/S_R$ is a decreasing function of the parameter $\beta$ indicating that the broader are the distribution of probability of duration of contacts the higher is the information encoded in the dynamics of the networks. Therefore the heterogeneity in the distribution of duration of contacts and no-interaction periods implies higher level of information in the social network. The human adaptive behavior by changing the exponent $\beta$ in face-to-face interactions and mobile phone communication effectively change the entropy of the dynamical network.

In conclusion, in the last ten years it has been recognized that the vast majority of complex systems can be described by networks of interacting units. Network theory has made tremendous progresses in this period and we have gained important insight into the microscopic properties of complex networks. Key statistical properties have been found to occur universally in the networks, such as the small world properties and broad degree distributions. Moreover the local structure of networks has been characterized by degree correlations, clustering coefficient, loop structure, cliques, motifs and communities. The level of information present in these characteristic of the network can be now studied with the tools of information theory.
An additional fundamental aspect of social networks is their dynamics. This dynamics encode for information and can be modulated by adaptive human behavior.
In this paper we have  introduced the entropy of social dynamical networks and we have evaluated the information present in dynamical data of phone-call communication.
By analysing the phone-call interaction networks we have shown that the entropy of the network depends on the circadian rhythms. 
Moreover we have shown that social networks are extremely adaptive and are modified by the use of technologies. The statistics of duration of phone-call indeed is described by a Weibull distribution that strongly differ from the distribution of face-to-face interactions in a conference. Finally we have evaluated how the information encoded in social dynamical networks change if we allow a parametrization of the duration of contacts mimicking the adaptability of human behavior.
Therefore  the entropy of social dynamical networks  is able to quantify   how the social networks dynamically change during the day and how they dynamically adapt to different technologies.

\section*{Material and Methods}
In order to describe the model of mobile phone communication, we consider a system consisting of $N$ agents  representing the mobile phone users. The agents are interacting in a  social network $G$ representing social ties  such as friendships, collaborations or acquaintances. The network  $G$ is weighted with the weights  indicating the strength of the social ties between agents.  We use $N_1^k(t_0,t)dt_0$ to denote the number of  agents with degree $k$ that at time $t$ are not interacting and have not interacted  with another agent since time $t'\in(t_0,t_0+{1}/{N})$. Similarly we  denote by $N_2^{k,k',w}(t_0,t)dt_0$ the number of connected agents  (with degree respectively $k$ and $k'$ and weight of the link $w$) that at time $t$ are interacting in phone call started at time $t'\in (t_0,t_0+1/N)$. The mean-field equation for this model read,
\begin{equation}
\frac{\partial N_1^k(t_0,t)}{\partial t}=-(1+ck){N_1^k(t_0,t)}f_1(t_0,t)+N\pi_{21}^k(t)\delta_{tt_0} \nonumber
\end{equation}
\begin{equation}
\frac{\partial N_2^{k,k',w}(t_0,t)}{\partial t}=-2N_2^{k,k',w}(t_0,t)f_2(t_0,t|w)+N\pi_{12}^{k,k',w}(t)\delta_{tt_0}
\label{dN1w}
\end{equation}
where the constant $c$ is given by 
\begin{equation}
c=\frac{\sum_{k'}\int_0^t dt_0 N_1^{k'}(t_0,t)f_1(t_0,t)}{\sum_{k'}k'\int_0^t dt_0N^{k'}_1(t_0,t)f_1(t_0,t)}.
\label{c_sum}
\end{equation}
 In Eqs. $(\ref{dN1w})$ the rates $\pi_{pq}(t)$  indicate the average number of agents changing from state $p=1,2$ to state $q=1,2$ at time $t$. These rates can be also expressed in a self-consistent way and the full system solved for any given choice of $f_1(t_0,t)$ and $f_2(t_0,t|w)$ (See Text S1 for details).

The definition of the  entropy of dynamical social networks of a pairwise communication model, is given by Eq. (6).
To evaluate the entropy of dynamical social network explicitly, we have to carry out the  summations in Eq. $(6)$. These sums, will in general depend on the particular history of the dynamical social network, but in the framework of the model we study, in the large network limit will be dominated by their average value. In the following therefore we perform these sum in the large network limit. The first summation in Eq. $(6)$ denotes the average loglikelihood of finding at time $t$ a  non-interacting agent  given a history ${\cal S}_t$. We can distinguish between two eventual situations occurring at  time $t$: {\it (i)} the agent has been  non-interacting since a  time $t-\tau$, and at time $t$ remains non-interacting; {\it (ii)} the agent has been interacting with another agent since time $t-\tau$, and at time $t$ the conversation is terminated by  one of the two interacting agents.The second term in the right hand side of  Eq. $(6)$, denotes the average loglikelihood of finding two agents in a connected pair at time $t$ given a history ${\cal S}_t$. There are two possible situations that might occur for two interacting agents  at time $t$: {\it (iii)} these two agents have been  non-interacting, and to time $t$ one of them  decides to form a connection with the other one; {\it (iv)} the two agents have been   interacting with each other since a time  $t-\tau$, and they remain interacting at time $t$. Taking into account all these possibilities we have been able to use the transition probability form different state and the number of agents in each state to evaluate the entropy of dynamical networks in the large network limit (For further details on the calculation see the Text S1).

 \section*{Ethics Statement}
The dataset used in this study only involved de-identified information and no details about the subscribers were made available to us.

\section*{Acknowledgments}
We thank A.-L. Barab\'asi for his useful comments and for the mobile call data used in this research. MK acknowledges the financial support from EU’s 7th Framework Program’s FET-Open to ICTeCollective project no. 238597

\onecolumngrid
\newpage

\begin{center}
\LARGE
Entropy of dynamical networks \\ Supporting Informations \\
\large
\vspace{.3in}
K. Zhao, M. Karsai and G. Bianconi
\vspace{.3in}
\end{center}

\normalsize
 
\maketitle

\section{The proposed model of cellphone communication }
\subsection{Dynamical social network for pairwise communication}
We consider a system consisting of $N$ agents  representing the mobile phone users. The agents are interacting in a  social network $G$ representing social ties  such as friendships, collaborations or acquaintances. The network  $G$ is weighted with the weights  indicating the strength of the social ties between agents. To model the mechanism of cellphone communication, the agents can call their neighbors in the social network $G$ forming  groups of interacting agents of size two. Since at any given time a call can be initiated or terminated the network is highly dynamical. We assign to each agent $i=1,2,\dots,N$ a coordination number $n_i$ to indicate his/her state. If $n_i=1$ the agent is non-interacting, and if $n_i=2$ the agent is in a mobile phone connection with another agent. The dynamical process of the model at each time step $t$ can be described explicitly by the following algorithm:
\begin{itemize}
\item[(1)] An agent $i$ is selected randomly at time $t$.
\item[(2)] 
The subsequent action of agent $i$  depends on his/her current state (i.e. $n_i$):
\begin{itemize}
\item[(i)] If $n_i=1$, he/she will call one of his/her non-interacting neighbors $j$ of $G$ with probability $f_1(t_i,t)$ where $t_i$ denotes the last time at which agent $i$ has changed his/her state. Once he/she decides to call, agent $j$ will be chosen randomly in between the neighbors of $i$ with probability proportional to $f_1(t_j,t)$, therefore  the coordination numbers of agent $i$ and $j$ are  updated according to the rule $n_i \rightarrow 2$ and $n_j \rightarrow 2$.
\item[(ii)] If $n_i=2$, he/she will terminate his/her current connection with probability $f_2(t_i,t|w_{ij})$ where $w_{ij}$ is the weight of the link between $i$ and the neighbor $j$ 
that is interacting with $i$. Once he/she decides to terminate the connection, the coordination numbers are then updated according to the rule $n_i \rightarrow 1$ and $n_j \rightarrow 1$.
\end{itemize}
\item[(3)] Time $t$ is updated as $t \rightarrow t+1/N$ (initially $t=0$) and the process is iterated until $t=T_{max}$. 
\end{itemize}

\subsection{General solution to the model }
In order to solve the model analytically, we assume the quenched network $G$ to be annealed and uncorrelated. Therefore we assume that at each time the network is rewired keeping the degree distribution $p(k)$ and the weight distribution $p(w)$ constant. Moreover we solve the model in the continuous time limit.Therefore we always approximate the sum over time-steps of size $\delta t=1/N$ by integrals over time. We use $N_1^k(t_0,t)dt_0$ to denote the number of  agents with degree $k$ that at time $t$ are not interacting and have not interacted  with another agent since time $t'\in(t_0,t_0+{1}/{N})$. Similarly we  denote by $N_2^{k,k',w}(t_0,t)dt_0$ the number of connected agents  (with degree respectively $k$ and $k'$ and weight of the link $w$) that at time $t$ are interacting in phone call started at time $t'\in (t_0,t_0+1/N)$.  Consistently with  the annealed approximation the probability that an agent with degree $k$ is called  is proportional to its degree. Therefore the rate equations of the model are given by 
\bea   
\frac{\partial N_1^k(t_0,t)}{\partial t}&=&-{N_1^k(t_0,t)}f_1(t_0,t)-ck{N_1^k(t_0,t)}f_1(t_0,t)+N\pi_{21}^k(t)\delta_{tt_0} \nonumber\\
\frac{\partial N_2^{k,k',w}(t_0,t)}{\partial t}&=&-2N_2^{k,k',w}(t_0,t)f_2(t_0,t|w)+N\pi_{12}^{k,k',w}(t)\delta_{tt_0}
\label{dN1w}
\eea
where the constant $c$ is given by 
\be
c=\frac{\sum_{k'}\int_0^t dt_0 N_1^{k'}(t_0,t)f_1(t_0,t)}{\sum_{k'}k'\int_0^t dt_0N^{k'}_1(t_0,t)f_1(t_0,t)}.
\label{c_sum}
\ee
 In Eqs. $(\ref{dN1w})$ the rates $\pi_{pq}(t)$  indicate the average number of agents changing from state $p=1,2$ to state $q=1,2$ at time $t$. These rates can be also expressed in a self-consistent way as
\bea
\pi_{21}^k(t)&=&\frac{2}{N}\sum_{k',w}\int_0^t dt_0 f_2(t_0,t|w)N_2^{k,k',w}(t_0,t)\nonumber\\
\pi_{12}^{k,k',w}(t)&=&\frac{P(w)}{CN}\int_0^t dt_0 \int_0^t dt_0'N_1^k(t_0,t) N_1^{k'}(t_0',t) f_1(t_0,t)f_1(t'_0,t)(k+k') 
\label{pi12k_sum}
\eea
where the constant $C$ is given by 
\be
C=\sum_{k'}\int_0^t dt_0 k'N^{k'}_1(t_0,t)f_1(t_0,t).
\label{C_sum}
\ee
The solution to Eqs.  (\ref{dN1w}) is given by 
\bea   
N_1^k(t_0,t)=N\pi_{21}^k(t_0)e^{-(1+ck)\int_{t_0}^t{f_1(t_0,t)}{}dt}\nonumber \\
N_2^{k,k',w}(t_0,t)=N\pi_{12}^{k,k',w}(t_0)e^{-2\int_{t_0}^t{f_2(t_0,t|w)}dt}
\label{gN2w}
\eea
which must satisfy the self-consistent constraints Eqs. (\ref{pi12k_sum}) and the conservation of the number of agents with different degree
\be 
\int dt_0 \big[N_1^k(t_0,t)+\sum_{k',w} N_2^{k,k',w}(t_0,t)\big]=Np(k).
\label{N_conserve} 
\ee
In the following we will  denote by $P^k_1(t_0,t)$  the probability distribution that an agent with degree $k$ is  non-interacting for a period from $t_0$ to $t$  and by $P^w_2(t_0,t)$  the probability that a connection of weight $w$ at time $t$ is active since time  $t_0$. It is immediate to see  that these distributions are given by the number of individual in a state $n=1,2$ multiplied by the probability of having a change of state, i.e.
\bea
P_1^k(t_0,t)&=&(1+ck)f_1(t_0,t)N_1^k(t_0,t)\nonumber \\
P_2^w(t_0,t)&=&2f_2(t_0,t|w)\sum_{k,k'}N_2^{k,k',w}(t_0,t). 
\label{P12}
\eea

\subsection{Stationary solution with specific $f_1(t_0,t)$ and $f_2(t_0,t)$} 
\label{stationary_sec}
In order to capture the behavior of the empirical data with a realistic   model, we have chosen  
\bea
f_1(t_0,t)&=&f_1(\tau)=\frac{b_1}{(1+\tau)^{\beta}}\nonumber \\
f_2(t_0,t|w)&=&f_2(\tau|w)=\frac{b_2g(w)}{(1+\tau)^{\beta}}
\label{f2t}
\eea
with parameters $b_1>0$, $b_2>0$, $0 \le\beta\le 1$ and arbitrary positive function $g(w)$. In  Eqs. $(\ref{f2t})$, $\tau$ is the duration time elapsed since the agent has changed    his/her state for the last time (i.e. $\tau=t-t_0$ ). The functions  of $f_1(\tau)$ and $f_2(\tau|w)$ are decreasing function of their argument $\tau$ reflecting the reinforcement dynamics  discussed in the main body of the paper. The function $g(w)$ is generally chosen as a decreasing function of $w$,  indicating that connected agents  with a stronger weight of link  interact typically for a longer time. We are especially interested in the  stationary state solution of the dynamics. In this regime we have that for large   times  $t\gg1$ the distribution of the number of agents is only dependent on $\tau$. Moreover the transition rates $\pi_{pq}(t)$ also converge to  a constant independent of $t$ in the stationary state. Therefore the solution of the stationary state will satisfy 
\bea
N^k_1(t_0,t)&=&N^k_1(\tau)\nonumber \\
N^{k,k',w}_2(t_0,t)&=&N^{k,k',w}_2(\tau)\nonumber\\
\pi_{pq}(t)&=&\pi_{pq}.
\label{spit}
\eea 
The necessary condition for the stationary solution to exist is that the summation of self-consistent constraints given by Eq. (\ref{c_sum}) and  Eq. (\ref{C_sum}) together with  the conservation law Eq. (\ref{N_conserve}) converge under  the  stationary assumptions Eqs. (\ref{spit}). The convergence depends on the value of the parameters $b_0$, $b_1$, $\beta$ and the choice of function $g(w)$. In particular, when $0 \le \beta <1$, the convergence is always satisfied. In the following subsections, we will characterize further the stationary state solution of this model in different limiting cases.

\subsubsection{Case $0<\beta<1$}
\label{beta_l1}
The expression for the number of agent in a given state $N_1^k(\tau)$ and $N_2^{k,k',w}(\tau)$ can be obtained by 
substituting Eqs. (\ref{f2t}) into the general solution Eqs. (\ref{gN2w}), using the stationary conditions Eqs. (\ref{spit}). In this way  we get the stationary solution given by  
\bea
 N_1^k(\tau)=N\pi_{21}^ke^{\frac{b_1(1+ck)}{1-\beta}[1-(1+\tau)^{1-\beta}]}=N\pi_{21}^km_1^k(\tau)\nonumber \\
 N_2^{k,k',w}(\tau)=N\pi_{12}^{k,k',w}e^{\frac{2b_2g(w)}{1-\beta}[1-(1+\tau)^{1-\beta}]}=N\pi_{12}^{k,k',w}m_2^w(\tau).
\label{N2w_tau}
\eea
To   complete the solution is necessary to determine the constants $\pi_{21}^k$ and $\pi_{12}^{k,k'w}$ in a self-consistent  type of solution.To find the expression of $\pi_{12}^{k,k',w}$ as a function of $\pi_{21}^k$ we substitute Eqs. $(\ref{N2w_tau})$ in  Eq.$(\ref{pi12k_sum})$ and we get 
\bea
\pi_{12}^{k,k',w}(t)&=&\frac{1}{C}\pi_{21}^kP(w)\bigg[k\int_0^t dt_0 m^k_1(t_0,t)f_1(t_0,t)\int_0^t dt_0'N^{k'}_1(t'_0,t)f_1(t'_0,t) \nonumber\\
&+&k'\int_0^t dt_0m^k_1(t_0,t)f_1(t_0,t) \int_0^t dt_0' N^{k'}_1(t'_0,t)f_1(t'_0,t)\bigg].
\label{pi12kkw}
\eea
Finally we get a closed equation for $\pi_{21}^k$ by substituting Eq.(\ref{pi12kkw}) in  Eq.(\ref{N_conserve})  and using the definition of $c$ and $C$, given respectively by Eq. $(\ref{c_sum})$ and Eq. $(\ref{C_sum})$. Therefore we get 
\bea
\pi_{21}^k&\bigg[&\int_0^{\infty}m^k_1(\tau)d\tau+\int_{w_{min}}^{w_{max}}P(w)\int_0^{\infty}m_2^w(\tau)d\tau dw \nonumber \\
&\times& \bigg(ck\int_0^{\infty}m_1^k(\tau)f_1(\tau)d\tau+\int_0^{\infty}m_1^k(\tau) f_1(\tau)d\tau \bigg)\bigg]=p(k). 
\eea
Performing explicitly the last two integrals using the dynamical solution given by Eqs. $(\ref{N2w_tau})$, this equation can be simplified as
\be
\pi_{21}^k=\bigg[\int_0^{\infty}m^k_1(\tau)d\tau+\int_{w_{min}}^{w_{max}}P(w)\int_0^{\infty}m_2^w(\tau)d\tau dw \bigg]^{-1}p(k).
\label{pi10k}
\ee
Finally the self-consistent solution of the dynamics is solved by expressing   Eq. $(\ref{c_sum})$  by 
\be
c=\frac{\sum_{k}\pi_{21}^k (1+ck)^{-1}}{\sum_{k} \pi_{21}^k k (1+ck)^{-1}}.
\label{cself}
\ee
Therefore we can use Eqs. $(\ref{pi10k})$ and $(\ref{cself})$ to compute the numerical value of $\pi_{21}^k$ and $c$. Inserting in these equations the expressions for $f_1(\tau),f_2(\tau|w)$ given by Eqs. $(\ref{f2t})$ and the solutions $N_1^k(\tau),N_2^{k,k',w}(\tau)$ given by Eqs. $(\ref{N2w_tau})$ we get
\bea
P_1^k(\tau)&\propto &\frac{b_1(1+ck)}{(1+\tau)^{\beta}}e^{-\frac{b_1(1+ck)}{1-\beta}(1+\tau)^{1-\beta}}\nonumber \\
P_2^w(\tau)&\propto &\frac{2b_2g(w)}{(1+\tau)^{\beta}}e^{-\frac{2b_2g(w)}{1-\beta}(1+\tau)^{1-\beta}}.
\label{P2kt}
\eea
The probability distributions $P^k_1(\tau)$ and $P^w_2(\tau)$, can be manipulating performing a data collapse of the distributions, i.e. 
\bea
\tau_1^{\star}(k)P_1^k\bigg(x_1=\frac{\tau}{\tau_1^{\star}(k)}\bigg)&=&A_1{x_1}^{-\beta}e^{-\frac{{x_1}^{1-\beta}}{1-\beta}}\nonumber \\
\tau_2^{\star}(w)P_2^w\bigg(x_2=\frac{\tau}{\tau_2^{\star}(w)}\bigg)&=&A_2{x_2}^{-\beta}e^{-\frac{{x_2}^{1-\beta}}{1-\beta}}
\label{transform2}
\eea
with $\tau_1^{\star}(k)$ and $\tau_2^{\star}(w)$ defined as 
\bea
\tau_1^{\star}(k)&=&\big[b_1(1+ck)\big]^{-\frac{1}{1-\beta}}\nonumber \\
\tau_2^{\star}(w)&=&\big[2b_2g(w)\big]^{-\frac{1}{1-\beta}}
\eea
where $A_1$ and $A_2$ are the normalization factors.  The data collapse defined by Eqs. (\ref{transform2}) of the curves $P_1^k(\tau)$, $P^w_2(\tau)$ and are  both described by Weibull distributions.
\subsection{Comparisons with quenched simulations}
\begin{figure*}[t]
\begin{center}
\includegraphics[width=150mm, height=50mm]{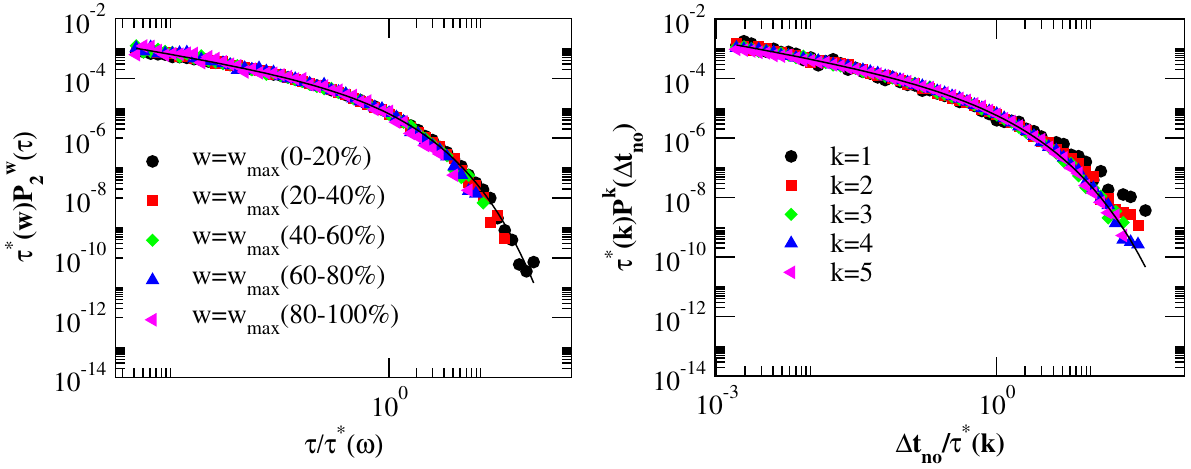}
\end{center}
\caption{Data collapse of the simulation of the proposed model for cell phone communication. In the  panel (A) we plot the probability  $P_2^w(\tau)$ that in the model a pair of agents with strenght $w$ are interacting for a period $\tau$ and in the  panel (B) we plot the probability  $P_1^k(\tau)$  that in the model an agents of degree $k$ is non-interacting for a period $\tau$  The simulation data on a quenched networks are compared with  the analytical predictions  (solid lines) in the annealed approximation. The collapses data of $P_2^w(\tau)$ is described by Weibull distribution in  agreement with the empirical results found in the mobile phone data.}
\label{sim}
\end{figure*}

To check the validity of our annealed approximation versus quenched simulations, we performed a computer simulation  according to the dynamical process  on a quenched network. In Fig. \ref{sim} we compare the results of the simulation  with the prediction of the analytical solution. In particular in the reported simulation we have  chosen $\beta=0.5$, $b_1=0.02$, $b_2=0.05$ and $g(w)=w^{-1}$, the simulation is based on a number of agent $N=2000$ and for a period of $T_{max}=10^5$, finally the data are averaged over $10$ realizations and the network is Poisson with average $\avg{k}=6$ and weight distribution $p(w)\propto w^{-2}$. In Fig. \ref{sim}, we show evidence that the Weibull distribution and the data collapse of $P^w_2(\tau)$ well capture the empirical behavior observed in the mobile phone data (Fig. \ref{interaction}). The distribution of the non-interaction periods $P_1^k(\tau)$ in the model is by construction unaffected by circadian rhythms but follow a similar data collapse as observed in the real data (Fig. \ref{noninteraction}). The simulated data are also in good agreement with the analytical prediction predicted in the annealed approximation for the parameter choosen in the figure. As the network becomes more busy and many agents are in a telephone call, the quenched simulation and the annealed prediction of $P_1^k(\tau)$ differs more significantly.

\subsubsection{Case $\beta=0$ }
  For $\beta =0$ the functions $f_1(\tau)$ and $f_2(\tau|w)$ given by Eqs.$(\ref{f2t})$ reduce to constants, therefore the process of creation of an interaction is a Poisson process and no reinforcement dynamics is taking place in the network. Assigning $\beta=0$ to Eqs. (\ref{gN2w}), we get the solution 
\bea
 N_1^k(\tau)&=&N\pi_{21}^ke^{-b_1(1+ck)\tau}\nonumber \\
 N_2^{k,k',w}(\tau)&=&N\pi_{12}^{k,k',w}e^{-2b_2g(w)\tau}.
\eea
and consequently the distributions of duration of given states Eqs. $(\ref{P12})$ are given by 
\bea
 P_1^k(\tau) \propto e^{-b_1(1+ck)\tau}\nonumber \\
P_2^w(\tau) \propto e^{-2b_2g(w)\tau}.
\eea
Therefore the  probability distributions $P_1^k(\tau)$  and $P_2^w(\tau)$  are  exponentials as expected in a Poisson process.

\subsubsection{Case $\beta=1$}
In this section, we discuss the case for $\beta=1$ such that $f_1^k(\tau)\propto(1+\tau)^{-1}$ and $f_2^w(\tau|w)\propto(1+\tau)^{-1}$. Using Eqs. (\ref{dN1w}) we get the solution 
\bea
 N_1^k(\tau)=N\pi_{21}^k(1+\tau)^{-b_1(1+ck)}\nonumber \\
 N_2^{k,k',w}(\tau)=N\pi_{12}^{k,k',w}(1+\tau)^{-2b_2g(w)}.
\eea
and consequently the distributions of duration of given states Eqs. $(\ref{P12})$ are given by 
\bea
 P_1^k(\tau) \propto \pi_{21}^k(1+\tau)^{-b_1(1+ck)-1}\nonumber \\
 P_2^w(\tau) \propto \pi_{12}^{k,k',w}(1+\tau)^{-2b_2g(w)-1}.
\eea
The probability distributions are power-laws.This result remains  valid for every value of the parameters $b_1,b_2,g(w)$ (See Ref. \cite{Zhao:2011} for a full account of the detailed solution of this model) nevertheless the stationary condition is only valid for  
\bea
b_1(1+ck)>1\nonumber \\
2b_2g(w)>1.
\eea
Indeed this condition ensures that the self-consistent constraits Eqs. (\ref{c_sum}), (\ref{C_sum}) and the conservation law Eq. (\ref{N_conserve}) have a stationary solution.

\subsection{Solution of the mean-field model on a fully connected network}
\label{mean_field_sec}
Finally, we discuss the mean-field limit on the model in which every agent can interact with every other agent. In this case, social network is a fully connected network. 
Therefore we use $N_1(t_0,t)$ and $N_2(t_0,t)$ to denote the number of agents of the two different states respectively and the rate equations are then revised to 
\bea   
\frac{\partial N_1(t_0,t)}{\partial t}&=&-2{N_1(t_0,t)}f_1(t_0,t)+N\pi_{21}(t)\delta_{tt_0} \nonumber \\
\frac{\partial N_2(t_0,t)}{\partial t}&=&-2{N_2(t_0,t)}f_2(t_0,t)+N\pi_{12}(t)\delta_{tt_0}
\eea
Since we will refer to this model only in the framework of a null model, we will only discuss the case in which the dynamics of the network is Poissonian, i.e. when
\bea 
f_1(t_0,t)&=&{b_1}\nonumber \\
f_2(t_0,t)&=&{b_2}.
\eea
The stationary solution of this model is given by  exponentials, i.e. 
\bea
 N_1(\tau)&=&N\pi_{21}e^{-2b_1\tau}\nonumber \\
 N_2(\tau)&=&N\pi_{12}e^{-2b_2\tau}.
 \label{mfe}
 \eea
Finally the distributions of duration of given states expressed by Eqs. $(\ref{P12})$ are given by 
\bea
P_1(\tau)&\propto &e^{-2b_1\tau} \nonumber \\
P_2(\tau)&\propto &e^{-2b_2\tau},
\eea
which are exponential distributions as expected in  a Poisson process.
\newpage
\section{Entropy of the dynamical social networks}

\subsection{Entropy of the dynamical social networks of pairwise communication}
\label{entropy_pairwise}

The definition of the  entropy of dynamical social networks of a pairwise communication model, is given by Eq. (6) of the main body of the article that we repeat here for convenience,
\bea
{ S}=&-&\sum_i P(g_i(t)=1|{\cal S}_t)\log P(g_i(t)=1|{\cal S}_t) \nonumber \\
&-&\sum_{ij}a_{ij} P(g_{ij}(t)=1|{\cal S}_t)\log P(g_{ij}(t)=1|{\cal S}_t)
\label{s_pair}
\eea
In this equation the matrix $a_{ij}$ is the adjacency matrix of the social network and $g_{ij}(t)=1$ indicates that at time $t$ the agents $i$ and $j$ are interacting while $g_i(t)=1$ indicates that agent $i$ is non-interacting.Finally ${\cal S}_t=\{g_{i}(t^{\prime}),g_{ij}(t^{\prime})\ \ \forall t^{\prime}<t\}$ indicates the dynamical evolution of the social network.
In this section, we will evaluate  the entropy of dynamical social networks in the framework of the annealed model of pairwise communication explained in detail in the previous section of this supplementary material.
To evaluate the entropy of dynamical social network explicitly, we have to carry out the  summations in Eq. $(\ref{s_pair})$. These sums, will in general depend on the particular history of the dynamical social network, but in the framework of the model we study, in the large network limit will be dominated by their average value. In the following therefore we perform these sum in the large network limit. The first summation in Eq. $(\ref{s_pair})$ denotes the average loglikelihood of finding at time $t$ a  non-interacting agent  given a history ${\cal S}_t$. We can distinguish between two eventual situations occurring at  time $t$: {\it (i)} the agent has been  non-interacting since a  time $t-\tau$, and at time $t$ remains non-interacting; {\it (ii)} the agent has been interacting with another agent since time $t-\tau$, and at time $t$ the conversation is terminated by  one of the two interacting agents. In order to characterize situation {\it (i)} we indicate by  $P_{1 \rightarrow 1}^k(\tau)$  the probability that a non-interacting agent with   degree $k$ in the social network, that  has not interacted since a  time $\tau$, doesn't change state. Similarly, in order to characterize situation {\it (ii)},  we indicate by  $P_{2 \rightarrow 1}^{k,k',w}(\tau)$ the probability that a connected pair of agents (with  degrees $k$ and $k'$ respectively, and weight of the link $w$) have interacted since time $\tau$ and terminate their conversation at time $t$. 
Given the stationary solution of the pairwise communication model, performed in the annealed approximation, the rates $P_{1 \rightarrow 1}^k(\tau)$ and  $P_{2 \rightarrow 1}^{k,k',w}(\tau)$ are given by
\bea
P_{1 \rightarrow 1}^k(\tau)&=&1-\frac{f_1(\tau)}{N}-\frac{kf_1(\tau)}{NC}\sum_{k'}\int N^{k'}_1(\tau')f_1(\tau')d\tau' \nonumber \\
&=& 1-(1+ck)\frac{f_1(\tau)}{N}\nonumber\\
P_{2 \rightarrow 1}^{k,k',w}(\tau)&=&\frac{2f_2(\tau|w)}{N}
\label{P21}
\eea
where the constant $C$ is given by
\be
C=\sum_{k'}\int k'N^{k'}_1(\tau')f_1(\tau')d\tau'        
\ee
and  $f_1(\tau)$ and $f_2(\tau|w)$ are given in Sec. \ref{stationary_sec}. The variable $N^k_1(\tau)$ indicates the number of  agents of connectivity $k$ noninteracting since a time $\tau$. This number can in general fluctuate but in the large network limit it converges to its mean-field value given by Eq. $(\ref{N2w_tau})$ The second term in the right hand side of  Eq. $(\ref{s_pair})$, denotes the average loglikelihood of finding two agents in a connected pair at time $t$ given a history ${\cal S}_t$. There are two possible situations that might occur for two interacting agents  at time $t$: {\it (iii)} these two agents have been  non-interacting, and to time $t$ one of them  decides to form a connection with the other one; {\it (iv)} the two agents have been   interacting with each other since a time  $t-\tau$, and they remain interacting at time $t$. To describe the   situation {\it (iii)}, we indicate by  $P_{1 \rightarrow 2}^{k,k'}(\tau,\tau')$ the probability that two non interacting agents, isolated since time $t-\tau$  and $t-\tau'$ respectively, interact at time $t$. In order to describe situation {\it (iv)}, we denote by  $P_{2 \rightarrow 2}^{k,k',w}(\tau)$ the probability that two interacting agents, in interaction since a time $t-\tau$, remain interacting at time $t$.
In the framework of the stationary annealead approximation of the dynamical network these probabilities are given by 
\bea
P_{1 \rightarrow 2}^{k,k'}(\tau,\tau')&=&\frac{f_1(\tau)f_1(\tau')}{NC}(k+k') \nonumber \\
P_{2 \rightarrow 2}^{k,k',w}(\tau)&=&1-\frac{2f_2(\tau|w)}{N}.
\label{P22}
\eea
Therefore, the entropy of dynamical social networks given by Eq. (\ref{s_pair}) can be evaluated in the thermodynamic limit, and in the annealed approximation, according to the expression
\bea
{\cal S}=&-&\sum_k\int_0^{\infty}N_1^k(\tau)P^k_{1\rightarrow1}(\tau)\log{P^k_{1\rightarrow1}(\tau)}d\tau \nonumber \\
&-&\sum_{k,k',w}\int_0^{\infty}N_2^{k,k',w}(\tau)P^{k,k',w}_{2\rightarrow1}(\tau)\log{P^{k,k',w}_{2\rightarrow1}(\tau)}d\tau \nonumber \\
&-&\frac{1}{2}\sum_{k,k'}\int_0^{\infty}\int_0^{\infty}N_1^k(\tau)N_1^{k'}(\tau')P^{k,k'}_{1\rightarrow2}(\tau,\tau')\log{P^{k,k'}_{1\rightarrow2}(\tau,\tau')}d\tau d\tau' \nonumber \\
&-&\frac{1}{2}\sum_{k,k',w}\int_0^{\infty}N_2^{k,k',w}(\tau)P^{k,k',w}_{2\rightarrow2}(\tau)\log{P^{k,k',w}_{2\rightarrow2}(\tau)}d\tau,
\label{s_pair2}
\eea
with $N_1^k(\tau) $ and $N_2^{k,k',w}(\tau)$ given in the large network limit by Eqs. $(\ref{N2w_tau})$.

\subsection{Entropy of the null model}
To understand the impact of the distribution of duration of the interactions and of the distribution of non-interaction periods, we have compared the entropy $S$ of the pairwise communication model   with the entropy $S_R$ of a null model. Here we use the exponential  mean-field model described in Section \ref{mean_field_sec} as our null model. In this model  the  agents are embedded in a fully connected networks and the probability of changing the agent state does not include the reinforcement dynamics. In fact we have that the transition rates are  independent of time ($\beta=0$) and given by  $f^R_1(\tau)=b_1^R$ and $f^R_2(\tau)=b_2^R$. Following the same steps used in Sec. \ref{entropy_pairwise} for  evaluating $S$ in the model of pairwise communication on the networks, it can be easily proved that the entropy  $S_R$  of the dynamical null model is given by  
\bea
S_R=&-&\int_0^{\infty}N^R_1(\tau)\bigg[1-\frac{2b_1^R}{N}\bigg]\log{\bigg[1-\frac{2b_1^R}{N}\bigg]}d\tau \nonumber \\
&-&\int_0^{\infty}N^R_2(\tau)\frac{2b_2^R}{N}\log{\frac{2b_2^R}{N}}d\tau \nonumber \\
&-&\frac{1}{2}\int_0^{\infty}\int_0^{\infty}N^R_1(\tau)N^R_1(\tau')\frac{2b_1^R}{NC^R}\log{\frac{2b_1^R}{NC^R}}d\tau d\tau' \nonumber \\
&-&\frac{1}{2}\int_0^{\infty}N_2^R(\tau)\bigg[1-\frac{2b_2^R}{N}\bigg]\log{\bigg[1-\frac{2b_2^R}{N}\bigg]}d\tau 
\eea
where the constant $C^R$ is given by 
\be
C^R=\int_0^{\infty} N_1^R(\tau)d\tau,
\ee
and where $N_1,N_2$ are given , in the large network limit by their mean-field value given by Eq.$(\ref{mfe})$.
In order to build an appropriate null model for the pairwise communication model parametrized by $(\beta, b_1, b_2)$ ,we take the  parameters of the null model  $b_1^R$ and $b_2^R$ such that the proportion of the total number of agents in the two states (interacting or non-interacting) is the same in the pairwise model of social communication and in the null model.
In order to ensure this condition we need to satisfy the following relation
\be
\frac{\sum_k\int_0^{\infty}N_1^k(\tau)d\tau}{\sum_{k,k',w}\int_0^{\infty}N_2^{k,k',w}(\tau)d\tau}=\frac{\int_0^{\infty}N_1^R(\tau)d\tau}{\int_0^{\infty}N_2^R(\tau)d\tau}.
\label{portion}
\ee
In particular we have chosen $b_1^R=b_1$ and we have used Eq. $(\ref{portion})$ to determine $b_2^R$.

\newpage
\section{Measurement of the entropy of a typical week-day of cell-phone communication from the data}
In this section we discuss the method of measuring the dynamical entropy from  empirical cellphone data as a function of time $t$ in a typical weekday. This analysis gave rise to the results presented in figure $2$ in the main body of the paper.
We have analyzed the call sequence of subscribers of a major European mobile service provider. We considered calls between users who at least once called each other during the examined $6$ months period in order to examine calls only reflecting trusted social interactions. The resulted event list consists of $633.986.311$ calls between $6.243.322$ users. For the entropy calculation we selected $562.337$ users who executed at least one call per a day during a working week period. Since the network is very large we have assumed that the dynamical entropy can be evaluate in the mean-field approximation.
 We measured the following quantities directly from the sample: 
 \begin{itemize}
 \item
  $N_1(\tau,t)$ the number of agents in the sample that at time $t$ are not in a conversation since time $t-\tau$; 
  \item $N^{calls}(\tau,t)$ the number of agents in the sample  that are not in a conversation since time $t-\tau$ and make a call at time $t$; 
  \item $N^{called}(\tau,t)$ the number of agents in the sample that are not in a conversation since time $t-\tau$ and are called at time $t$; 
  \item $M^{in}(\tau,t)$ the number of agents that at time $t$ are in a conversation of duration $\tau$ with another agent in the  sample; 
  \item  $M^{out}(\tau,t)$ the number of agents that at time $t$  are in a conversation of duration $\tau$ with another agent outside  the sample;
  \item $M^{end}(\tau,t)$ the number of calls of duration $\tau$ that end at time $t$. 
  \end{itemize}Using the above quantities, we estimated the probability $p^{calls}(\tau,t)$ that an agent  makes a call at time $t$ after a non-interaction period of duration $\tau$, the probability $p^{called}(\tau,t)$ that an agent is called at time $t$ after a non-interaction period of duration $\tau$ and the probability $\pi(\tau,t)$ that a call of duration $\tau$ ends at time $t$,according to the following relations
\bea
p^{calls}(\tau,t)&=&\frac{N^{calls}(\tau,t)}{N_1(\tau,t)} \nonumber \\
p^{called}(\tau,t)&=&\frac{N^{called}(\tau,t)}{N_1(\tau,t)} \nonumber \\
\pi(\tau,t)&=&\frac{M^{end}(\tau,t)}{M^{in}(\tau,t)/2+M^{out}(\tau,t)}.
\eea
Since the sample  of $562.337$ users we are considering is a subnetwork of the whole dataset constituted by $6.243.322$ users, 
in our measurement, an agent can be  in one of three possible states
\begin{itemize} \item {\it state 1:} the agent is non-interacting; \item {\it state 2:} the agent is in a conversation with another agent of the sample; \item{\it state 3:} the agent is in a conversation with an agent outside the sample.
\end{itemize} Therefore , to evaluate the entropy of the data, we can modify Eq.(\ref{s_pair}) into
\bea
{\cal S}(t)=&-&\sum_i P(g_i(t)=1|{\cal S}_t)\log P(g_i(t)=1|{\cal S}_t) \nonumber \\
&-&\sum_{ij}a_{ij} P(g_{ij}(t)=1|{\cal S}_t)\log P(g_{ij}(t)=1|{\cal S}_t) \nonumber \\
&-&\sum_i P(g'_i(t)=1|{\cal S}_t)\log P(g'_i(t)=1|{\cal S}_t)
\label{s_data}
\eea
where $a_{ij}$ is the adjacency matrix of the quenched social network,  $g_i(t)=1$ indicates that the agent $i$ is in state 1, $g_{ij}(t)=1$ indicates that the agent is in state 2 interacting with agent $j$ and  $g'_i(t)=1$ indicates the agent $i$ is in state 3. Finally ${\cal S}_t=\{g_{i}(t^{\prime}),g_{ij}(t^{\prime})\ g'_i(t)\ \ \forall t^{\prime}<t\}$ indicates the dynamical evolution of the social network. To explicitly evaluate Eq. (\ref{s_data}) in the large network limit where we assume that the dependence on the particular history are vanishing, we sum over the loglikelihood of all transitions between different states using the same strategy in Sec.2, which is 
\bea
{\cal S}(t)=&-&\sum_{\tau}N_1(\tau,t)P_{1\rightarrow1}(\tau,t)\log{P_{1\rightarrow1}(\tau,t)} \nonumber \\
&-&\sum_{\tau}M^{in}(\tau,t)P_{2\rightarrow1}(\tau,t)\log{P_{2\rightarrow1}(\tau,t)} \nonumber \\
&-&\sum_{\tau}M^{out}(\tau,t)P_{3\rightarrow1}(\tau,t)\log{P_{3\rightarrow1}(\tau,t)} \nonumber \\
&-&\frac{1}{2}\sum_{\tau,\tau'}N_1(\tau,t)N_1(\tau',t)P_{1\rightarrow2}(\tau,\tau',t)\log{P_{1\rightarrow2}(\tau,\tau',t)} \nonumber \\
&-&\frac{1}{2}\sum_{\tau}M^{in}(\tau,t)P_{2\rightarrow2}(\tau,t)\log{P_{2\rightarrow2}(\tau,t)} \nonumber \\
&-&\sum_{\tau}N_1(\tau,t)P_{1\rightarrow3}(\tau,t)\log{P_{1\rightarrow3}(\tau,t)} \nonumber \\
&-&\sum_{\tau}M^{out}(\tau,t)P_{3\rightarrow3}(\tau,t)\log{P_{3\rightarrow3}(\tau,t)}.
\label{s_data2}
\eea
where the probabilities of transitions between different states are given by
\bea
P_{1\rightarrow1}(\tau,t)&=&1-p^{calls}(\tau,t)-p^{called}(\tau,t) \nonumber \\
P_{2\rightarrow1}(\tau,t)&=&P_{3\rightarrow1}(\tau,t)=\pi(\tau,t) \nonumber \\
P_{1\rightarrow2}(\tau,\tau',t)&=&\frac{(1-\gamma)}{C}\bigg[p^{calls}(\tau,t)p^{called}(\tau',t)+p^{calls}(\tau',t)p^{called}(\tau,t)\bigg] \nonumber \\
P_{2\rightarrow2}(\tau,t)&=&P_{3\rightarrow3}(\tau,t)=1-\pi(\tau,t) \nonumber \\
P_{1\rightarrow3}(\tau,t)&=&\gamma\bigg[p^{calls}(\tau,t)+p^{called}(\tau,t)\bigg]
\label{P_transition}
\eea
and where $C$ is given by 
\be
C=\sum_{\tau} N_1(\tau,t)p^{called}(\tau,t).
\ee
Finally in $\ref{P_transition}$ we have introduced a parameter $\gamma \in[0,1]$ to denote the portion of the calls occurring between an agent in the sample and an agent out of the sample. For simplicity, we assume that $\gamma$ is a constant.  Substituting Eq.(\ref{P_transition}) into Eq.(\ref{s_data2}), we have performed the  summation over $\tau$ to obtain the value of entropy as a function of $t$ presented in Figure 2 of the main body of the paper where we have taken $\gamma=0.8$, consistently with the data.

\end{document}